\documentclass[sigconf]{webmedia}
\usepackage{comment}

\AtBeginDocument{%
  \providecommand\BibTeX{{%
    \normalfont B\kern-0.5em{\scshape i\kern-0.25em b}\kern-0.8em\TeX}}}

\setevent{webmedia}
\proceedingsDetails[WebMedia’2024]{Proceedings of the Brazilian Symposium on Multimedia and the Web}{2024}{Juiz de Fora, Brazil} 
\ISSN{2966-2753}

\begin{document}

\title[A critical analysis of Fediverse decentralization promises]{An evidence-based and critical analysis of the Fediverse decentralization promises}
%\subtitle{Subtitle (optional)}
%% Of note is the shared affiliation of the first two authors and the
\author{Henrique S. Xavier}
\email{hxavier@nic.br}
\affiliation{%
  \institution{Núcleo de Informação e Coordenação do Ponto BR}
  \streetaddress{Av. das Nações Unidas, 11541, 7º andar}
  \city{São Paulo}
  \state{SP}
  \postcode{04578-000}
}

\renewcommand{\shortauthors}{Xavier, H. S.}

\begin{abstract}
This paper examines the potential of the Fediverse, a federated network of social media and content platforms, to counter the centralization and dominance of commercial platforms on the social Web. We gather evidence from the technology powering the Fediverse (especially the ActivityPub protocol), current statistical data regarding Fediverse user distribution over instances, and the status of two older, similar, decentralized technologies: e-mail and the Web. Our findings suggest that Fediverse will face significant challenges in fulfilling its decentralization promises, potentially hindering its ability to positively impact the social Web on a large scale.        
\end{abstract}


\begin{comment}
\begin{teaserfigure}
  \includegraphics[width=\textwidth]{images/fediverse_teaser}
  \caption{Representation of federated instances of Wordpress, Mastodon, Funkwhale, Pixelfed and PeerTube.}
  \Description{A graph of logos of federated open-source social media and content platforms.}
  \label{fig:teaser}
\end{teaserfigure}
\end{comment}

\maketitle

\section{Introduction}
\label{sec:introduction}

\begin{figure}[h]
  \centering
  \includegraphics[width=\linewidth]{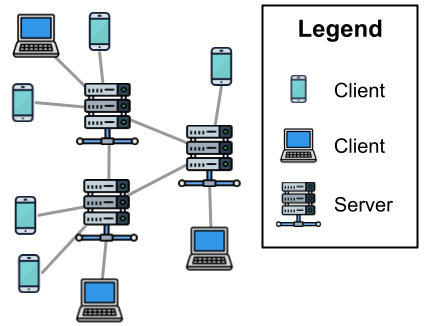}
  \caption{Representation of a federated system: clients only connect to a single server (instance), which in turn communicates with other servers.}
  \Description{Undirected graph in which cellphones and laptops, representing clients, connect to servers. These servers connect to one another.}
  \label{fig:federation-graph}
\end{figure}

The Fediverse is a network of servers hosting social media platforms where each server is known as an instance and where users on one instance can interact with users on different instances. As shown in Fig. \ref{fig:federation-graph}, the users do not interact directly as they would on a peer-to-peer network. Instead, a user employs client software (e.g., a web browser) to connect, post, and retrieve content from the instance on which he/she has an account. This instance, in turn, may send and retrieve user content to/from other instances by employing an agreed-upon inter-instance communication protocol. Note that the instances do not share a copy of all the content in the Fediverse. Alternatively, instance $A$ only receives content from a user on instance $B$ if $B$ is not moderated (i.e., blocked) by the instance $A$ and if the user in $B$ is followed by some user on $A$.   

Apart from the standard inter-instance communication protocol, each instance can be entirely independent regarding ownership, infrastructure, and software. This independence can lead to varying mechanics, designs, and approaches to governance, moderation, content recommendation, and data collection practices. Hence, this network of social media is referred to as ``federated:'' it is composed of autonomous instances that communicate with each other.    

Among the various existing communication protocols, such as OStatus, Diaspora, Matrix, Nostr, and Bluesky's AT\footnote{https://www.w3.org/community/ostatus, https://diasporafoundation.org, https://matrix.org, https://nostr.com, https://atproto.com}, ActivityPub \cite{Lemmer-Webber2018} -- recommended by the World Wide Web Consortium (W3C) -- is currently the most widely used. It is employed by Mastodon, the most pervasive social media software in the Fediverse, in terms of both user numbers\footnote{This fact excludes proprietary platforms like Threads that also implement the ActivityPub protocol.} and instances, as reported by various Fediverse observatories.\footnote{https://fedidb.org, https://fediverse.party, https://fediverse.observer, https://instances.social} 

While most instances in the Fediverse operate on free and open-source software (FOSS) like Mastodon, Friendica, Pixelfed, PeerTube and Lemmy\footnote{https://joinmastodon.org, https://friendi.ca, https://pixelfed.org, https://joinpeertube.org, https://join-lemmy.org}, there are also proprietary platforms that can communicate with others (thus participating in the Fediverse) through the ActivityPub protocol, such as Threads and Flipboard\footnote{https://threads.net, https://flipboard.com}. Despite ActivityPub being implemented in some proprietary platforms, many Fediverse practitioners consider only FOSS to be integral to the Fediverse (e.g., the Fediverse observatories mentioned above).

The Fediverse has generated excitement among tech enthusiasts and internet users with its potential to challenge the dominance of Big Tech social media platforms. It promises to make the Web a decentralized environment once again and to create safer, healthier, and more enjoyable social media experiences, with less hate speech, disinformation, advertising, censorship, and algorithmic manipulation \cite{Alquimidia2024,Cantrill2022,Davie2023,Pierce2024}. Typically, the rationale is that the inter-instance communication protocol would ensure decentralization, preventing user lock-in effects. The decentralization would result in a wide variety of instances for users to choose from -- many run by enthusiasts and communities not driven by profit. By selecting the ``best'' instances, users would theoretically make the Fediverse a better environment than current large platforms.

The promise mentioned above, however, is not based on direct empirical evidence but instead deduced from a theoretical social model that includes assumptions about the mechanisms governing interactions between technology, society, and individuals. This paper aims to critically and evidence-basedly assess the decentralization claim and its consequences by scrutinizing the Fediverse's technology, examining current characteristics of the Fediverse itself, and comparing it with the history of similar, more mature systems. Specifically, we want to verify what the existing evidence suggests about the Fediverse as it gains scale regarding the following research questions:

\begin{enumerate}
\item Will fediverse users be dispersed on many instances (i.e. decentralized)? \label{rq:descentralization} 
\item Will non-profit organizations replace big techs as the main players on the Fediverse? \label{rq:for-profit}
\item Does the federation architecture contribute to solving current issues like the spread of hate speech and disinformation? \label{rq:issues}
\end{enumerate}
Our analysis will focus on the ActivityPub protocol and the social media platforms connected by it.  
\section{Analysis}

\subsection{Technological details}
\label{sec:tech}

The first step in our analysis was to double-check common claims made about the Fediverse using the documentation of the software associated with the Fediverse as a primary source, along with direct tests of the software and detailed accounts from practitioners regarding specific issues. Through this process, we linked these claims to specific parts of the software and the system's architecture and verified if the technical specifications directly support them or if the claims rely on social mechanisms beyond the technology employed.

It is essential to distinguish between social media and platform FOSS from the inter-instance communication protocol and its implementation. The former facilitates the creation of individualized social media instances (e.g., instances of Mastodon or Peertube), allowing control over governance, funding policies, behavioral rules, and moderation practices. Moreover, modifying the source code provides further control over design, content recommendation, privacy, and data collection practices.

On the other hand, the inter-instance communication protocol primarily enables federation, standardizing communication between instances.\footnote{ActivityPub, in particular, also standardizes the communication between client and server software \cite{Lemmer-Webber2018}, thus enabling the development of multiple universal clients that can communicate with several servers.} In summary, the inner workings of each instance are determined by the FOSS (and potential modifications thereof), not by the inter-instance communication protocol responsible for federating the instances. As we will explore, many of the Fediverse's characteristics stem directly from the vanilla versions of FOSS code rather than from the federated nature enabled by the protocol.

The claim that users can easily switch between instances, taking their data, posts, followers, and followings with them \cite{Pierce2024, Davie2023}, is contradicted by the documentation of Mastodon\footnote{https://docs.joinmastodon.org/user/moving/\#move} -- the most common instance type -- and by practitioners \cite{Rolltime2022}. While it is possible to migrate profiles and followers, migrating posts is not supported. Additionally, the ability to migrate followers may vary depending on the implementation running on the instances. Therefore, even if we discount customized instances that might restrict migration or discourage it through attractive features, some degree of out-of-the-box lock-in effect already exists in Mastodon.

% Fazer o teste acima.

There are often claims that, in the Fediverse, one can post to any instance from a single account \cite{Pierce2024}. However, strictly speaking, this is not accurate. To post on a server, one must have an account on that specific server (see Fig. \ref{fig:federation-graph}), and a post made on one instance $A$ will only reach another instance $B$ if the $A$ is not blocked by $B$ and if someone in $B$ follows the author of the post \cite{Lemmer-Webber2018}. Even though instances exchange content in certain situations, posting on an instance using an account from a different instance is analogous to posting on X (former Twitter) using your account on Instagram: it is impossible. Furthermore, content may not display correctly when shared between instances of different types (e.g., Lemmy and Funkwhale\footnote{https://www.funkwhale.audio}). These limitations may necessitate users to maintain accounts on multiple instances.

% Verificar se Lemmy vs funkwhale é um bom exemplo.

The claim that Fediverse instances are controlled by communities rather than corporations, while not disproven by its technologies as seen in the cases above, is not directly supported by them either. In other words, it is not a direct consequence of the technical properties of ActivityPub or the code of social media FOSS. On the other hand, some claims do align with the vanilla versions of FOSS: timelines display content from one's followers without advertisements, sponsored content, or algorithmic recommendations, and user data is not repurposed for tasks such as AI training. However, no technical or legal barriers prevent modifications to FOSS code, and ActivityPub does not inherently block instances from the Fediverse that implement data usage, recommendation algorithms, or sponsored content. Additionally, nothing prevents proprietary social media platforms with arbitrary internal operations and implementations of ActivityPub from participating in the Fediverse. 

Finally, the claim that the Fediverse prevents or reduces censorship has a more direct connection to Fediverse technology. The ability of instances to communicate with each other and the ease of setting up a social media FOSS instance and creating user accounts promotes a less controllable environment without single points of failure.

\subsection{Current Fediverse characteristics}
\label{sec:fediverse}

The fediverse technology indeed enables the formation of a network comprising independent social media instances capable of communicating with each other. However, this decentralized structure introduces a discoverability challenge \cite{Pierce2024}, as there is no central or unified index where all instances and users can be found. To identify existing instances and their characteristics, such as the number of active users, fediverse observatories utilize crawlers. These crawlers begin with known instances and trace connections to other instances through user-following relationships. In our study, we used data from \texttt{fediverse.observer}, collected on August 2, 2024, to estimate the distribution of users across various instances. 

\begin{figure}[h]
  \centering
  \includegraphics[width=\linewidth]{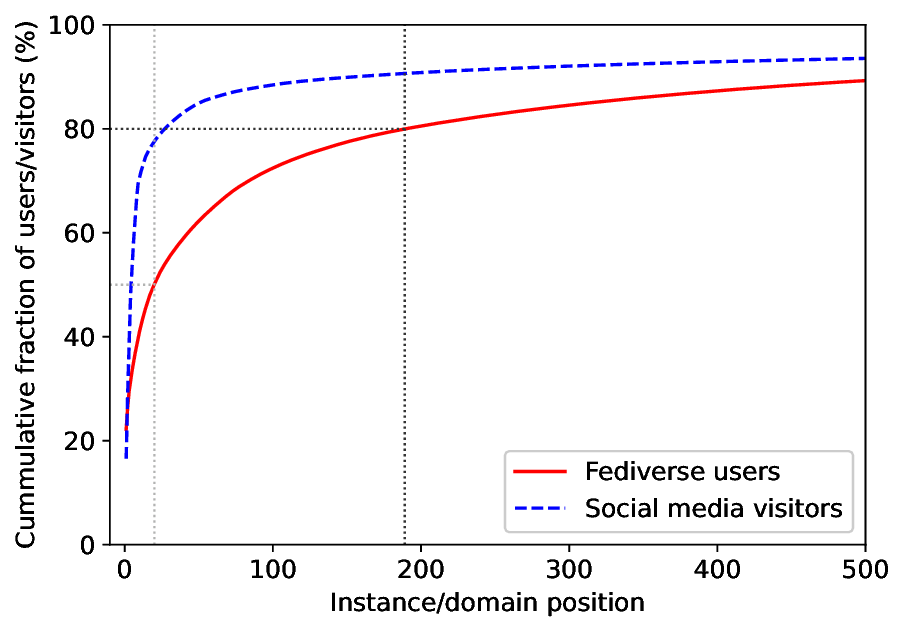}
  \caption{The solid red (dashed blue) line depicts the cumulative fraction of monthly active users (unique visitors) on the top fediverse instances (social media domains).}
  \Description{Both curves show fast growth initially and flatten out as more instances are included. The curve for unique visitors to social media domains is steeper initially, indicating that this distribution is more concentrated than the Fediverse monthly active users.}
  \label{fig:fediverse-concentration}
\end{figure}

Although the communication protocol enables decentralization within the Fediverse, not all instances hold equal prominence. Our data reveals that while large instances like \texttt{mastodon.social} have approximately 225,000 monthly active users as of August 2024, 82\% of the 21,146 instances in our dataset have fewer than five active users, and 24\% have no active users at all. The solid red line in Fig. \ref{fig:fediverse-concentration} represents the cumulative fraction of monthly active users concentrated on the top $k$ instances. The gray dotted lines show that 50\% of all active users are concentrated on the top 20 instances, and 80\% of all active users are found on 189 instances, which constitutes only 0.9\% of the instances in our dataset. 

Fig. \ref{fig:fediverse-concentration} also shows, as a dashed blue line, the cumulative fraction of unique monthly visitors to the top 10,000 domains classified as ``Social Networks and Online Communities'', according to web traffic data from SimilarWeb \cite{Xavier2024}. Besides top fediverse instance domains, this dataset also includes non-federated social media platforms like \texttt{facebook.com}, \texttt{tiktok.com}, \texttt{linkedin.com}, \texttt{reddit.com}, \texttt{ok.ru}, and \texttt{tumblr.com}. While unique visitors is not the same as active users (users must login on the instance, while unique visitors do not), the concepts are close enough to indicate that the concentration on social media in general is larger than currently in the Fediverse.

Our data also shows that the FOSS-based Fediverse has approximately one million active users. In contrast, Meta's Threads boasted 150 million active users as of April 2024 \cite{Mehta2024}. Threads' dominance within the ActivityPub ecosystem is further evident in terms of the most popular user accounts: as of June 2024, the four accounts with the highest number of followers on the Fediverse were hosted on Threads (Barack Obama with 4.5 million, Mark Zuckerberg with 4.1 million, Joe Biden with 3 million, and the White House with 1 million followers)\footnote{https://archive.ph/wip/NRFU8}. Therefore, the data indicates that Fediverse users are concentrated on a small number of instances and that most users are currently on a Big Tech for-profit platform. Other for-profit entities making their way into the Fediverse include Flipboard\footnote{https://archive.ph/NZsC3} \cite{Quagliarello2024}, WordPress \cite{Pfefferle2024}, and Bluesky (albeit with a distinct protocol). It is important to note that currently, Fediverse users can follow Threads accounts, but Threads users cannot follow or see content from other instances \cite{Su2024}. This asymmetry makes Threads only partially connected (one-way) to the Fediverse and illustrates the varied implementations of ActivityPub.

Website traffic data collected and provided by SimilarWeb also indicates that Threads and gab.com -- a for-profit alt-right social media platform \cite{Than2023} developed from a fork of Mastodon\footnote{https://github.com/GabOpenSource} -- had a higher average number of monthly visits between March and May 2024 compared to mastodon.social. Table \ref{tab:traffic} presents a comparison of their traffic alongside that of baraag.net, a Mastodon instance known for allowing cartoon child pornography.

\begin{table}
  \caption{Average traffic to selected instances connected to the Fediverse (Mar--May 2024).}
  \label{tab:traffic}
  \begin{tabular}{lr}
    \toprule
    Domain & Avg. monthly visits (millions)\\
    \midrule
    threads.net     &  303.4 \\
    gab.com         &    5.0 \\
    baraag.net      &    2.9 \\
    mastodon.social &    2.6 \\
  \bottomrule
\end{tabular}
\end{table}

Regarding Fediverse claims beyond decentralization, a casual exploration of the instances reveals a wide diversity in design, content, moderation, and behavior rules. Many inspected instances lack advertisements, sponsored content, or recommendation algorithms. However, some instances do exhibit these features. For example, mastodon.social, the most prominent Mastodon instance, offers (simple) recommendations in its ``explore'' section, while gab.com includes sponsored content. Gab also adopts a one-way approach to ActivityPub, preventing its users from following external accounts.

\subsection{Lessons from other decentralized systems}
\label{sec:other-systems}

Examining the development of older decentralized technologies like email and the Web can help anticipate the fate of the Fediverse. Email has a structure very similar to that of the Fediverse: it is based on a shared communication protocol, and in principle, anyone can set up an independent email server. Moreover, ActivityPub can be understood as ``JSON over email.'' All inter-instance communication is handled by sending JSON-LD-formatted text data from a user's outbox (hosted at the user's instance) to another user's inbox at a different instance \cite{Lemmer-Webber2018}.

According to data from SimilarWeb, traffic to webmail services is highly concentrated. Among the 2,406 domains classified by SimilarWeb under the "Email" industry, leaders like live.com and mail.ru accounted for 44.8\% and 25.9\% of the industry's global traffic between August and October 2023. It is worth noting that other significant webmail domains, such as mail.google.com, are not included in SimilarWeb's list because they are subdomains \cite{SimilarWeb2024b}. While live.com recorded a monthly average of 2.19 billion visits during this period, mail.google.com had 10.88 billion visits. Therefore, webmail services may be more concentrated than reported by SimilarWeb. 

Even when email domains remain distinct and diverse, the actual provision of email services may become concentrated. Liu \emph{et al.} demonstrated that organizations increasingly outsourced their email services, with 28.5\% and 10.8\% of the most popular domains (selected from Alexa's top one million domains between 2017 and 2021) having their email services provided by Google and Microsoft, respectively. Among a random sample of \texttt{.com} domains, primarily comprised of less popular domains, GoDaddy serves as the email service provider for 29\% of them \cite{Liu2021}. Practices implemented by large providers, such as shadow banning IP addresses purportedly to combat spam, may contribute to this concentration \cite{Fenollosa2022}.

Another decentralized technology example is the World Wide Web, an information system where clients retrieve content from independent servers using a straightforward public-domain protocol (HTTP) \cite{BernersLee1990,CERN1993}. Despite the simplicity of setting up a web server and the existence of over 100 million web domains \cite{SimilarWeb2024c}, traffic to these domains is highly concentrated: approximately 3,000 domains reportedly accumulate 50\% of global web traffic \cite{Xavier2024}. Among the 10,000 domains categorized by SimilarWeb as ``Social Networks and Online Communities'' --  one of the top 5 industries in terms of web traffic out of 187 -- these domains collectively received 50 billion monthly visits between August and October 2023. This industry demonstrates significant concentration, with Meta alone capturing 52\% of the industry's traffic during this period by aggregating traffic from facebook.com, instagram.com, and whatsapp.com. This concentration issue is precisely what the Fediverse aims to address, making its decentralization approach akin to the concept portrayed in the movie ``Inception''\footnote{https://www.imdb.com/title/tt1375666/}. SimilarWeb's data include the most visited Fediverse instances, with mastodon.social ranking 282nd in the industry's list, accounting for less than 0.01\% of its traffic.

\section{Discussion}

The Fediverse is often touted as a means to (re)decentralize the Web and counter the oligopoly dominating social media and content platforms \cite{Zignani2018,Davie2023,Alquimidia2024,Pierce2024}. Some of these advocates seem to argue that the inter-instance communication protocol itself promotes decentralization by empowering users to choose instances that align best with their preferences \cite{Mansoux2020}. As stated in \cite{Struett2023}, the Fediverse ``is often cast as inherently democratizing because of its uniquely federated architecture''. Conversely, some scholars adopt a more critical perspective by highlighting the challenges the Fediverse may encounter \cite{Struett2023} or by raising potentially problematic social impacts it could have \cite{Allen2023,Mahadeva2024}.

Regarding our research questions \ref{rq:descentralization} and  \ref{rq:for-profit}, we showed in Sec. \ref{sec:tech} that the Fediverse technology, on its own, does not prevent its potential capture by large for-profit companies or the concentration of users and instances among a few dominant players. In Sec. \ref{sec:fediverse}, empirical data revealed that users are heavily concentrated on a small number of instances, and for-profit entities, including Big Tech companies, are exerting significant influence. Lastly, Sec. \ref{sec:other-systems} examines how older decentralized systems such as email and the Web evolved into centralized structures dominated by a few significant players.

Our preliminary analysis indicates that while the Fediverse has the potential to reduce (though not eliminate) centralization compared to non-federated social media (see Fig. \ref{fig:fediverse-concentration}) and can foster communities tailored to the interests of the individual -- such as \texttt{eorzea.photos} for Final Fantasy players, \texttt{liberdon.com} for libertarians, \texttt{lemmygrad.ml} for Marxists, and \texttt{mexico-pensante.blog} for public policy discussions in Mexico -- it faces significant challenges in achieving a more even distribution of users and traffic across the social Web on a large scale (see Table \ref{tab:traffic} for an example). In essence, the Fediverse may address the needs of specific individuals and provide suitable local environments, but it is unlikely to disrupt, by itself, the dominance of for-profit web oligopolies.

Among the challenges faced by the Fediverse, as discussed in the literature, techno-romanticism may not only hinder its ability to achieve its goals but also help explain the gap between the overconfidence of some advocates and the evidence presented in this work. According to Struett \emph{et al.}, techno-romanticism involves three tropes \cite{Struett2023}:

\begin{enumerate}
\item The belief that technologies inherently possess the ability to improve the human condition regardless of social and political stewardship \cite{Streeter1987}.
\item The notion that individuals can significantly impact society without the support of institutions or collective coordination \cite{Streeter2010}.
\item The idea that free-market solutions are viable, often disregarding pre-existing power imbalances, and will inevitably increase social utility.
\end{enumerate}
Additional challenges to scaling the non-commercial Fediverse include financial sustainability, the risk of commercial capture by Big Tech, the maintenance of FOSS, and effective content moderation \cite{Struett2023}. Addressing these challenges will require moving beyond the aforementioned techno-romantic tropes, seeking institutional support, coordinating large-scale actions, confronting power imbalances, and pursuing evidence-based solutions. Practical steps might involve passing legislation to regulate large, for-profit social media platforms and creating incentives for federated alternatives; providing state funding for Fediverse infrastructure, maintenance, moderation, and FOSS development; and encouraging large, influential non-commercial institutions to adopt and promote the use of federated social media.

Beyond analyzing whether the Fediverse can disrupt Big Tech monopolies on the Web and how to circumvent the challenges it might face, it is crucial to gather evidence of the potential impacts of the Fediverse on pressing issues currently affecting the Web. These include the spread of hate speech and disinformation \cite{Shin2018,CastanoPulgarin2021}, effects on mental health \cite{Haidt2024}, and manipulation of user behavior \cite{Zuboff2019}. The presence of large instances like Gab, which permit hate speech, demonstrates that the Fediverse is not immune to some of these problems. In fact, its decentralized nature could complicate moderation efforts and potentially exacerbate some of these issues \cite{Allen2023,Mahadeva2024}.

In future developments of this study, we aim to delve deeper into the possibilities and limitations of ActivityPub and other federating protocols like AT, considering their selective implementation and potential vulnerabilities to manipulation. While social media FOSS also significantly influences the Fediverse, its openness introduces considerable variability that complicates forecasting its consequences. Additionally, we plan to survey existing instances, analyzing their specific characteristics and user distribution firsthand rather than relying on third-party analyses. Furthermore, we intend to engage extensively with existing literature to strengthen connections and debates on this topic. Ultimately, our goal is to link the anticipated evolution of the Fediverse to its potential consequences over the social challenges we currently face on the Web.  

%%%% ANOTAÇÕES %%%% 

% Truthsocial is based on Mastodon but defederated.

% Estatísticas de número de usuários/instâncias por software 

% Big Techs com práticas ruins tomaram o espaço de outras redes anteriores (e.g. MySpace, Orkut, Tiktok), mesmo com o vendor lock in. Sem vendor lock in, possivelmente elas tomariam os espaços do mesmo jeito.

% ``advertising-based models demand engagement — and algorithms will naturally observe that when we are enraged, we stay engaged. If follows that when algorithms are promoting content that enrages us, the social network will — either implicitly or explicitly — select for those that are most divisive.'' \cite{Cantrill2022}

%%%% FIM DAS ANOTAÇÕES %%%%

%\begin{acks}
%To Robert, for the bagels and explaining CMYK and color spaces.
%\end{acks}

\balance

\bibliographystyle{ACM-Reference-Format}
\bibliography{main}

%\appendix


\end{document}